\documentclass[twocolumn,showpacs,preprintnumbers,prl,fleqn]{revtex4}

\usepackage{graphicx}
\def\ul{  } 
\begin{document}
\title{
Gate-controlled spin-splitting in quantum dots with ferromagnetic
leads \\ in the Kondo regime }
\author{J.~Martinek,$^{1,3,6}$ M.~Sindel,$^2$ L.~Borda,$^{2,4}$
 J.~Barna\'s,$^{3,5}$ R.~Bulla,$^{7}$ J.~K\"onig,$^{8}$
 G.~Sch\"on,$^1$ S.~Maekawa,$^{6}$ and J.~von~Delft$^2$}
 \affiliation{$^1$Institut f\"ur Theoretische
Festk\"orperphysik, Universit\"at Karlsruhe, 76128 Karlsruhe,
Germany\\
 $^2$
Physics Department and Center for NanoScience, LMU M\"unchen,
80333 M\"unchen, Germany \\
 $^3$Institute of Molecular
Physics, Polish Academy of Sciences, 60-179 Pozna\'n, Poland \\
 $^4$ Institute of Physics, TU Budapest, H-1521, Hungary \\
  $^5$Department of Physics, Adam Mickiewicz University, 61-614 Pozna\'n,
 Poland \\
 $^6$Institute for Materials Research, Tohoku University, Sendai
980-8577, Japan \\
 $^7$Theoretische Physik III, Elektronische Korrelationen und
    Magnetismus, Universit\"at Augsburg, Augsburg, Germany \\
 $^8$Institut f\"ur Theoretische Physik III, Ruhr-Universit\"at Bochum, 44780
Bochum, Germany }

\date{\today}

 \begin{abstract}

The effect of a gate voltage ($V_g$) on the spin-splitting of an
electronic level in a quantum dot (QD) attached to ferromagnetic
leads is studied in the Kondo regime using a generalized numerical
renormalization group technique. We find that the $V_g$-dependence
of the QD level spin-splitting strongly depends on the shape of
the density of states (DOS). For one class of DOS shapes there is
nearly no $V_g$-dependence, for another, $V_g$ can be used to
control the magnitude and sign of the spin-splitting, which can be
interpreted as a local exchange magnetic field. We find that the
spin-splitting acquires a new type of logarithmic divergence. We
give an analytical explanation for our numerical results and
explain how they arise due to spin-dependent charge fluctuations.

\end{abstract}



\maketitle

 The manipulation of magnetization and spin
is one of the fundamental processes in magneto-electronics and
spintronics, providing the possibility of writing information in a
magnetic memory \cite{maekawa}, and also because of the
possibility of classical or quantum computation using spin.
 In most situations this is realized by means of an
externally applied, nonlocal magnetic field which is usually
difficult to insert into an integrated circuit. Recently, it was
proposed to control the magnetic properties, such as the Curie
temperature 
 of ferromagnetic
semiconductors, by means of an electric field: In gated structures
\cite{ohno}, due to the modification of carrier-density-mediated
magnetic interactions, such properties can be modified by a gate
voltage.
 In this Letter we propose to
control the amplitude and sign of the spin-splitting of a quantum
dot (QD) induced by the presence of ferromagnetic leads, only by
using a gate voltage without further assistance of a magnetic
field. To illustrate this effect we investigate the Kondo effect
and its spin-splitting as a very sensitive probe of the spin state
of the dot and the effective local magnetic field in the QD
generated by exchange interaction with the ferromagnetic leads.

Recently, the possibility of the Kondo effect in a QD attached to
ferromagnetic electrodes was widely discussed
\cite{sergueev,zhang,bulka,lopez,martinek1,martinek2,choi}, and it
was shown, that the Kondo resonance is split and suppressed in the
presence of ferromagnetic leads \cite{martinek1,martinek2}. It was
shown that this splitting can be compensated by an appropriately
tuned external magnetic field, and the Kondo effect is thereby
restored \cite{martinek1,martinek2}. In all previous studies of
QDs attached to ferromagnetic leads
\cite{sergueev,zhang,bulka,lopez,martinek1,martinek2,choi} an
idealized, flat, spin-independent DOS with spin-dependent
tunneling amplitudes was considered.
 However, since the
spin-splitting arises from renormalization effects i.e. is a
many-body effect, it depends on the {\it full} DOS-structure of
the involved material, and not only on its value at
  the Fermi surface.
 In realistic ferromagnetic
systems, the DOS shape is strongly asymmetric due to the Stoner
splitting and the different hybridization between the electronic
bands \cite{maekawa}.

In this Letter we {\ul demonstrate} that the gate voltage
dependence of the spin-splitting of a QD level, resulting in a
splitting and suppression of the Kondo resonance, {\ul is
determined by the DOS structure and can lead to crucially
different behaviours.}
 We apply
the numerical renormalization group (NRG) technique extended to
handle bands of arbitrary shape.
 For one class of DOS-shapes, we find almost no $V_g$-dependence of the spin-splitting, while for another class the
induced spin-splitting, which can be interpreted as the effect of
a local exchange field, can be controlled by $V_g$. The
spin-splitting can be fully compensated and its direction can even
be reversed within this class. We explain the physical mechanism
 that leads to this behavior, which is related to
the compensation of the renormalization of the spin-dependent QD
levels induced by the electron-like and hole-like quantum charge
fluctuations. Moreover we find that
 for the QD level close to the Fermi surface,
the amplitude of the
spin-splitting has a logarithmic divergence, indicating the
 many-body character of this phenomenon.

{\it Model and method}. --
 The Anderson model (AM) of a single level QD with energy $\epsilon_0$ and Coulomb interaction $ U $,
coupled to ferromagnetic leads, is given by
\vspace{-0.6mm}
\begin{eqnarray}
H &=& \sum_{rk \sigma} \epsilon_{rk \sigma}c_{rk\sigma}^{\dagger}
  c_{rk \sigma}
  + \epsilon_0 \sum_{\sigma} \hat n_\sigma + U
  \hat n_\uparrow \hat n_\downarrow
\nonumber\\
  && + \sum_{r k \sigma} (V_{r k} d_{\sigma}^{\dagger}
    c_{r k \sigma} + h.c.)
   - B S_{z}
  \; .
  \label{eq:AMf}
\end{eqnarray}
Here $c_{rk\sigma}$ and $d_\sigma$ ($  \hat n_\sigma =
d_{\sigma}^{\dagger} d_{\sigma} $) are Fermi operators for
electrons with momentum $k$ and spin $\sigma$ in the leads
($r=L/R$), and in the QD, $V_{rk}$ is the tunneling amplitude,
$S_{z} = (\hat n_\uparrow- \hat n_\downarrow)/2$, and the last
term denotes the Zeeman energy of the dot. The energy $\epsilon_0$
is experimentally controllable by $V_g$ ($\epsilon_0 \simeq V_g$).

In order to discuss the gate voltage dependence of the QD level
spin-splitting, we consider here a more realistic, both energy and
spin dependent band structure [$ \rho_{r \uparrow}(\omega) \neq
\rho_{r \downarrow}(\omega) $], violating p-h symmetry $ \rho_{r
\sigma}(\omega) \neq \rho_{r \sigma}(-\omega) $, which leads to an
energy dependent hybridization function $ \Gamma_{r \sigma} (
\omega ) = \pi \sum_{ k } \delta ( \omega - \epsilon_{k \sigma} )
V_{r k}^2 = \pi \rho_{r \sigma}(\omega) V^2_0 $, where we take $
V_{r k} = V_r $ to be constant.
We apply the NRG method \cite{costi1,hewson-book} extended to
handle arbitrary DOS shapes and asymmetry. To this end, the
standard logarithmic discretization of the conduction band is
performed for {\it each} spin component separately, with the
bandwidths, $D_\uparrow = D_\downarrow = D_0$, chosen such that
the total spectral weight is included in $[-D_0,D_0]$ for all
values of $V_g$ studied here (to avoid different systematic errors
upon changing $V_g$).

Within each interval $[-\omega_n,-\omega_{n+1}]$ and
$[\omega_{n+1},\omega_{n}]$ (with $\omega_n = D_0 \Lambda^{-n}$)
of the logarithmically discretized conduction band (CB)
 the operators of the continuous CB are
expressed in terms of a Fourier series.
 Even though we allow
for a non-constant conduction electron DOS, it is still possible
to transform the Hamiltonian such that the impurity couples {\it
only} to the zeroth order component of the Fourier expansion of
each interval \cite{bulla}.
Dropping the non-constant Fourier-components of each interval
 \cite{costi1,hewson-book} then results in a discretized version of the Anderson model
with the continous spectrum in each interval replaced by a single
fermionic degree of freedom (independently for both spin
directions).
Since we allow for an arbitrary DOS for {\it each} spin component
$\sigma$ ($\uparrow,\downarrow$) of the CB this mapping needs to
be performed for each $\sigma$ separately. This leads to the
Hamiltonian:
\begin{eqnarray}
  &H& =   \sum_{\sigma} \epsilon_{\sigma} \hat n_\sigma + U
  \hat n_\uparrow \hat n_\downarrow
   + \sqrt{ {\xi_{0\sigma}}/{\pi}}\sum_{\sigma} [
  d^\dagger_{\sigma}f_{0\sigma} +
   f^\dagger_{0\sigma}d_{\sigma}  ]  \nonumber \\
   & +& \sum_{\sigma n=0}^\infty [
   \varepsilon_{n\sigma} f_{n\sigma}^\dagger f_{n\sigma}
+ t_{n\sigma} ( f_{n\sigma}^\dagger f_{n+1\sigma}
 + f_{n+1\sigma}^\dagger f_{n\sigma})] \; ,
  \label{eq:H_si}
\end{eqnarray}
where $ f_{n \sigma} $ are fermionic operators at the $n$th site
of the Wilson chain, $ \xi_{0 \sigma} =1/2 \int_{-D_0}^{+D_0}
\Gamma_{\sigma}(\omega) d\omega $, $ t_{n \sigma} $ denotes the
hopping matrix elements, and $\epsilon_{\sigma}=\epsilon_0 - B
S_z$.
%
The absence of particle-hole symmetry leads to the appearance of
non-zero on-site energies, $\varepsilon_{n\sigma}$ along the
chain.
%
In this {\it general} case no closed expression for the matrix
elements $t_{n\sigma}$ and $\varepsilon_{n\sigma}$, both depending
on the particular structure of the DOS via
$\Gamma_{\sigma}(\omega)$, is known, therefore they have to be
determined recursively. This requires rather advanced numerical
methods, due to the exponentially fast decay of $ t_{n \sigma} $
and $\varepsilon_{n\sigma}$ along the chain \cite{tong}.

%

%
\begin{figure}[t!]
\centerline{\includegraphics[width=1\linewidth]{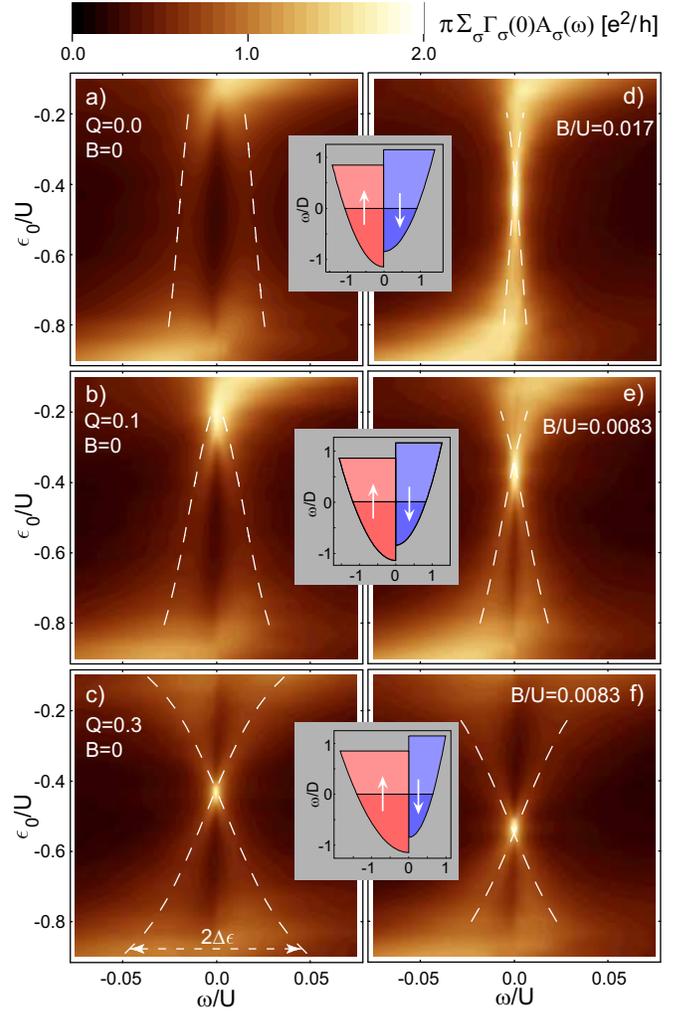}}
\caption{(color online). {\it $V_g$-dependence of the
spin-splitting:} Normalized spectral function $ \pi \sum_{\sigma}
\Gamma_{\sigma}(0) A_{\sigma}(\omega)  $ as a function of energy $
\omega $ and gate voltage $ \epsilon_0 $,  for the three different
DOS shapes (depicted in insets) characterized by a different $Q$,
{\ul which modifies both the spin and p-h asymmetry}: (a-c) for
magnetic filed $ B = 0 $, (d) $ B/U =0.017 $, (e), and (f) $ B/U =
0.0083 $. The white dashed lines are obtained using
Eq.~(\ref{eq:split_int}). Here $ U = 0.12 D_0 $, $ \pi V_0^2 = U
D/6 $, $ \Delta =0.15 D $ and $ T = 0 $.
 Inset: the scheme of the parabolic DOS
shape for spin $ \uparrow $ (red) and $ \downarrow $ (blue).
}
 \label{fig1}
\end{figure}

This method allows one to calculate the level occupation $
n_\sigma \! \equiv \! \langle \hat n_\sigma\rangle $ and the
spin-resolved single-particle spectral density $ A_{
\sigma}(\omega)  = -\frac{1}{\pi} {\rm Im} {\cal
G}_{\sigma}^{\mathrm r} (\omega)$, where ${\cal
G}_{\sigma}^{\mathrm r} (\omega)$ denotes a retarded Green's
function. For symmetric coupling [$ \Gamma_{L \sigma }(\omega) =
\Gamma_{R \sigma }( \omega)$] the spin-resolved conductance takes
the form $G_{\sigma} = \pi {e^2 \over h }
 \int_{- \infty } ^{ + \infty} d \omega \Gamma_{\sigma}( \omega)
 A_{\sigma}(\omega) (-{ \partial f( \omega ) \over
\partial \omega } ) $
where $f( \omega )$ is the Fermi function.

{\it Spectral function and conductance}. -- Here, we focus our
attention on $T=0$ properties.
 We have
analyzed several types of DOS shapes and found three typical
classes of the $V_g$-dependence of the Kondo resonance splitting.
Since our method enables us to perform NRG calculations for
arbitrary band-shapes, we decide to choose an example which turns
out to encompass all three classes, namely $\rho_{\sigma}(\omega)=
\frac{1}{2}{3 \sqrt{2} \over 8} D^{-3/2} (1+\sigma Q)\sqrt{\omega
+ D + \sigma \Delta} $, where $\omega \in [-D-\sigma \Delta ,D
-\sigma \Delta]$, $ D_0 = D + \Delta $, [$\sigma \equiv 1(-1) $
for $ \uparrow(\downarrow) $], a square-root shape DOS equivalent
to a parabolic band (as for free electrons) with Stoner splitting
$ \Delta$ \cite{yosida}, and some additional spin asymmetry $ Q $,
which modifies the amplitude of the DOS [see
Fig.~\ref{fig1}(insets)].

 In Fig.~\ref{fig1} we present the weighted spectral function
$ \tilde{A}(\omega) \equiv \pi {e^2 / h} \sum_{\sigma}
\Gamma_{\sigma}(0) A_{\sigma}(\omega) $, normalized such that for
$ \omega=0 $ it corresponds to the linear conductance $ G =
\tilde{A}(0) $, as a function of energy $ \omega $ and $
\epsilon_0 $. 
 We focus on a narrow energy window
around the Fermi surface where the Kondo resonance appears; charge
resonances are visible when $ \epsilon_0 $ or $ U + \epsilon_0 $
approach the Fermi surface, namely at energies $ \epsilon_0/U
\gtrsim -0.1 $ or $\lesssim -0.9 $.
Although the NRG method is designed to calculate equilibrium
transport, one can still roughly deduce, from the spin-splitting
of the Kondo resonance of the equilibrium spectral function $
\tilde{A}(\omega) $, the splitting of the zero-bias anomaly $
\Delta V $ in the non-equilibrium conductance $ G(V) $, since $
e\Delta V \sim 2 \Delta \epsilon $ \cite{martinek1} ($ \Delta
\epsilon \equiv \tilde{\epsilon}_{\uparrow} -
\tilde{\epsilon}_{\downarrow} $ is the splitting of the
renormalized levels).

We present $ \tilde{A}(\omega) $ for three DOS shapes depicted in
the insets of Fig.~\ref{fig1}: (i) $Q=0$ (a,d), (ii) $Q=0.1$
(b,e), and (iii) $Q=0.3$ (c,f), with $ 2 \Delta = 0.3 D $
\cite{papa}, leading to the three typical behaviors.
Here the parameter $Q$ tunes the spin and p-h asymmetry [see the
definition of $\rho_\sigma(\omega)$] resulting in different
behaviours (for a detailed discussion see the last section).
For (i) we find nearly any $ \epsilon_0 $-dependence of the
spin-splitting; for (ii), a strong $ \epsilon_0 $-dependence
without compensation of the spin-splitting (i.e. no crossing), and
for (iii) a strong $ \epsilon_0 $-dependence with a compensation
(i.e. a crossing) and a change of the direction of the QD
magnetization. The compensation (crossing) corresponds to the very
peculiar situation where the Kondo effect (strong coupling fixed
point) can be recovered in the presence of ferromagnetic leads
without any external magnetic field. A behavior as presented in
Fig.~\ref{fig1}(a,b) was recently observed experimentally
\cite{nygard,ralph}, where indeed a variation of the gate voltage
results in two split conduction lines $ G(V,V_g) $ which are
parallel for one case and converging for the other case, similar
to our findings.

\begin{figure}[t!]
\centerline{\includegraphics[width=1\linewidth]{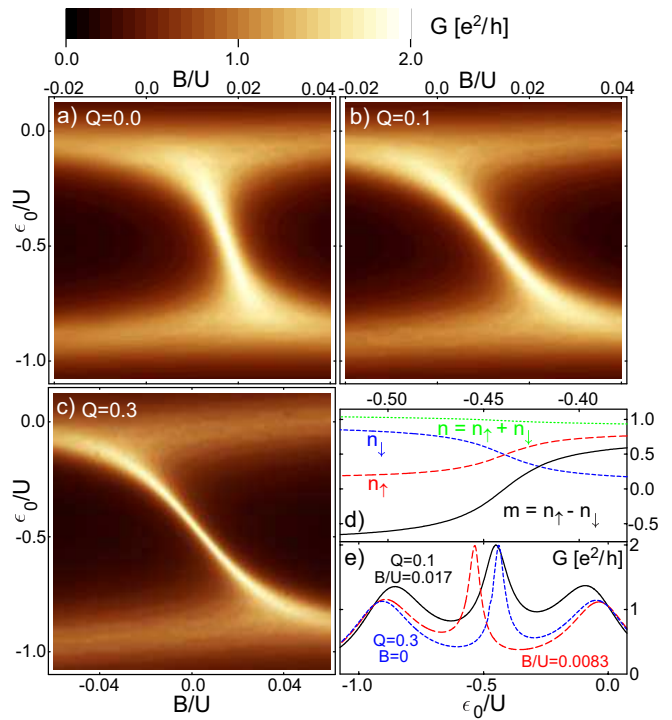}}
\caption{(color online). The QD's linear conductance $ G $ as a
function of gate voltage $ \epsilon_0 $ and external magnetic
field $ B $ for the DOS shapes (a), (b), and (c) as for
Fig.~\ref{fig1}(a), (b), and (c) respectively.
 (d) Spin-dependent occupancy $ n_{\sigma} $ of the dot level as a
function of gate voltage $ \epsilon_0 $ for the DOS shape as in
Fig.~\ref{fig1}(c) and $ B = 0 $. (d) The  $\epsilon_0 $-dependence
of the total occupancy of the dot $ n $
and magnetization $ m $ for the situation from Fig.~\ref{fig1}(c).
(e) The conductance G for the situations from Fig.~\ref{fig1}(c - dashed),
(d - solid), and (f - long dashed).
 Parameters $ U $, $ \Gamma $, and $ T $ as in Fig.~\ref{fig1}. }
 \label{fig2}
\end{figure}

{\it Effect of a magnetic field}. --
In Fig.~\ref{fig1}(d,e,f) we show how a magnetic field $B$
modifies the results of Fig.~\ref{fig1}(a,b,c): in (i) the spin
splitting can be compensated at a particular magnetic field $
B_{\mathrm{ comp }}$ (here $B_{\mathrm{ comp }}/U = 0.017 $) and
the Kondo effect is visible in a wide range of $\epsilon_0$; for
(ii), at $ B/U = 0.0083 $, the Kondo effect is
 recovered only at one particular $\epsilon_0$-value, which
depends on the applied magnetic field; case (iii) shows that the
crossing point shifts with $ B $.
 Since $ B_{\mathrm{ comp }}$
can be viewed as a measure of the zero-field splitting, $\Delta
\epsilon(B=0,\epsilon_0) \simeq - B_{\mathrm{ comp
}}(\epsilon_0)$, the $\epsilon_0$-dependence of $\Delta \epsilon$
 can be measured
by studying that of $B_{\mathrm{ comp }}$, for which one needs to
measure the linear conductance $G(\epsilon_0,B)$ as a function of
both $B$ and $\epsilon_0$.
In Fig.~\ref{fig2}(a-c) we plot
 $G(\epsilon_0,B)$ for the three bands of Fig.~\ref{fig1}.
  The two horizontal ridges (resonances) in
Fig.~\ref{fig2}(a-c) correspond to quantum charge fluctuations
(broadened QD level) of width $ \sim \Gamma $.
The lines with finite slope in Fig.~\ref{fig2}(a-c) reflect the
restored Kondo resonance and hence map out the
$\epsilon_0$-dependence of $B_{\mathrm{ comp }}(\epsilon_0) = -
\Delta \epsilon(\epsilon_0)$ when the magnetic field compensates
the spin-splitting.
Interestingly the spin-splitting and the corresponding
$B_{\mathrm{ comp }}$ tend to diverge ($ | \Delta \epsilon |
\rightarrow \infty $) when approaching the charging resonance, as
is best visible in Fig.~\ref{fig2}(c).

Such a finite slope in $ G(\epsilon_0,B)$ was observed for a
singlet-triplet transition Kondo effect in a two level QD
(Fig.2(d) Ref.~\cite{sasaki}). The corresponding transition leads
to a characteristic maximum in the valley between two charging
resonances (Fig.3(c) Ref.~\cite{sasaki}), similarly as in our
Fig.~\ref{fig2}(e). In that system the effective spin asymmetry
(assumed by our model) is realized by the asymmetry in the
coupling of two QD levels \cite{boese}.

In Fig.~\ref{fig2}(d) we show how the occupation $ n_{\sigma }$
and the magnetic moment (spin) of the QD $ m = n_{\uparrow} -
n_{\downarrow} = 2 \langle S_z \rangle $ change as a function
of $ \epsilon_0 $ for the situation of Fig.~\ref{fig1}(c). One
finds that even though $ B = 0 $, it is possible to control the
level spin-splitting of the
 QD, i.e. its spin, and thereby change the average spin direction
of the QD from the parallel to anti-parallel
alignment w.r.t. the lead's magnetization. This opens the
possibility of controlling the QD's spin state by means of a gate
voltage without further need of an external magnetic field, which
is difficult to apply locally in practical devices.

{\it Perturbative analysis}. --
 One can understand the
behavior presented in Fig.~\ref{fig1}(a-c) by using Haldane's
scaling method \cite{haldane}, where charge fluctuations are
integrated out. This leads to a spin-dependent renormalization of
the QD's level position $ \tilde{\epsilon}_{\sigma}$ and a level
broadening $\Gamma_{\sigma}$. In contrast to Ref.~\cite{martinek1}
we consider here the case of finite Coulomb interactions $ U <
\infty $, which means that also the doubly occupied state $ |2
\rangle $ is of importance. The spin-splitting is then given by $
\Delta \epsilon \equiv \delta\epsilon_{\uparrow} -
\delta\epsilon_{\downarrow} + B $, where
\begin{eqnarray}
\delta\epsilon_{\sigma} \simeq -\frac{1}{\pi} \int d\omega \left\{
\frac{\Gamma_{\sigma}(\omega)[1-f(\omega)]}{\omega -
\epsilon_{\sigma}} +
\frac{\Gamma_{-\sigma}(\omega)f(\omega)}{\epsilon_{-\sigma}+U-\omega}
 \right\} \, .
  \label{eq:split_int}
\end{eqnarray}
The first term in the curly brackets corresponds to electron-like
processes, namely charge fluctuations between a single occupied
state $ |\sigma \rangle $ and the empty $ |0 \rangle $ one, and
the second term to hole-like processes, namely charge fluctuations
between the states $ |\sigma \rangle $ and $ |2 \rangle $.
The amplitude of the charge fluctuations is proportional to $
\Gamma $, which for $ \Gamma \! \gg \!\! T $ determines the width
of QD's levels.
Eq.~(\ref{eq:split_int}) shows that $ \Delta \epsilon $ depends on
the shape of $ \Gamma_{\sigma}(\omega) $ for all $\omega$, not
only on its value at the Fermi surface.
The dashed lines in  Fig.~\ref{fig1}(a-c) show $\pm \Delta
\epsilon $ as a function of $\epsilon_0$ [from
Eq.~(\ref{eq:split_int})] for the same set of parameters as in the
NRG calculation, and are in good agreement with the position of
the (split) Kondo resonances observed in the latter.
Eq.~(\ref{eq:split_int}) shows that the dramatic changes observed
in Fig.~\ref{fig1} upon changing $Q$ are due to the modification
of the p-h and spin asymmetry.

Eq.~(\ref{eq:split_int}) predicts that even for systems with
spin-asymmetric bands $ \Gamma_{\uparrow}(\omega) \neq
\Gamma_{\downarrow}(\omega) $, the integral can give $ \Delta
\epsilon = 0 $, which corresponds to a situation where the
renormalization of $ \epsilon_{\sigma} $ due to electron-like
processes are compensated by hole-like processes. An example is a
system consisting of p-h symmetric bands, $ \Gamma_{ \sigma
}(\omega) = \Gamma_{ \sigma }(- \omega )$, for which there is no
splitting of the Kondo resonance ($ \Delta \epsilon = 0 $) for the
symmetric point, $ \epsilon_0 = - U/2 $.
For real systems p-h symmetric bands cannot be assumed, however
the compensation  $ \Delta \epsilon = 0 $ is still possible, as in
Fig.~\ref{fig1}(c).
Eq.~(\ref{eq:split_int}) also shows that the characteristic energy
scale of the spin-splitting is given by $ \Gamma $ rather than by
the Stoner splitting $ \Delta $ ($ \Delta \gg \Gamma $), since the
states far from the Fermi surface enter Eq.~(\ref{eq:split_int})
only with a logarithmic weight. However, the Stoner splitting
introduces a strong p-h asymmetry, so it {\ul can} influence the
character of gate voltage dependence significantly.

 For a
flat band $ \Gamma_{\sigma}( \omega ) = \Gamma_{\sigma} $,
Eq.~(\ref{eq:split_int}) can be integrated analytically. For $ D_0
\gg U$, $ |\varepsilon_0 | $ one finds:
$ \Delta \epsilon \simeq \left(  P \; \Gamma / \pi\right)\;
\mathrm{Re} [ \;\phi(\epsilon_0) - \phi(U+ \epsilon_0) \; ] $,
where $ P \equiv ( \Gamma_{\uparrow} - \Gamma_{\downarrow}
)/\Gamma $, $ \phi(x) \equiv \Psi ( \frac{1}{2} + i \frac{x}{2 \pi
T} ) $, and $\Psi(x)$ denotes the digamma function. For $ T = 0 $,
the spin-splitting is given by
\begin{eqnarray}
\Delta \epsilon \simeq \left( P \; \Gamma \;/\pi\right)
 \ln (  {| \epsilon_0 |}/{| U + \epsilon_0 |} )  \; ,
  \label{eq:split_T0}
\end{eqnarray}
showing a logarithmic divergence for $ \epsilon_0 \rightarrow 0 $
or $ U + \epsilon_0 \rightarrow 0$.
Since any sufficiently smooth DOS can be
 linearized around the Fermi surface, this logarithmic
divergence occurs quite universally, as can be observed in
log-linear versions (not shown) of Fig.~\ref{fig2}(a-c).
For finite temperature ($ T>0 $) the logarithmic divergence for $
\epsilon_0\rightarrow 0 $ or $\epsilon_0 \rightarrow -U$ is cut
off, $ \Delta \epsilon \simeq - {1 \over
 \pi} P \Gamma [ \Psi( {1 \over 2} ) + \ln {2 \pi T \over U} ] $,
which is also important for temperatures $ T \ll
T_{\mathrm K} $.

In conclusion, we demonstrated, using the extended NRG technique
for general band shapes, the possibility of controlling the local
exchange field and thereby the spin-splitting in a QD attached to
ferromagnetic leads by means of the gate voltage. A new type of
the logarithmic divergence
of the QD's level spin-splitting was found,
and attributed to spin-dependent charge fluctuations.

We thank T. Costi, L. Glazman, W. Hofstetter, B. Jones, C. Marcus,
J. Nyg{\aa}rd, A. Pasupathy, D. Ralph, A. Rosch, M. Vojta, and Y.
Utsumi for discussions. This work was supported by the DFG under
the CFN, 'Spintronics' RT Network of the EC RTN2-2001-00440, and
Project OTKA T034243.

\end{document}